\documentclass[conference]{IEEEtran}
\IEEEoverridecommandlockouts

\usepackage{cite}
\usepackage{amsmath,amssymb,amsfonts}
\usepackage{graphicx}
\usepackage{textcomp}
\usepackage{xcolor}
\usepackage{array}
\usepackage{tabularx}
\usepackage{url}
\usepackage{pgfplots}
\pgfplotsset{compat=newest}
\usepackage[strings]{underscore}
\pgfplotsset{compat=1.18}
\usepgfplotslibrary{polar}
\usepgfplotslibrary{statistics}
\usepackage{algorithm}
\usepackage{algpseudocode}
\usepackage{pgfplots}
\pgfplotsset{compat=1.18}
\usetikzlibrary{arrows.meta}
\usepackage{booktabs}
\usetikzlibrary{arrows.meta,positioning,calc}
\usepackage{makecell}
\definecolor{colBlue}{HTML}{0072B2}
\definecolor{colVermillion}{HTML}{D55E00}
\definecolor{colGreen}{HTML}{009E73}
\definecolor{colOrange}{HTML}{E69F00}
\definecolor{colPurple}{HTML}{CC79A7}
\usepackage{tikz}
\usetikzlibrary{arrows.meta, positioning,  shapes.geometric, calc}

\def\BibTeX{{\rm B\kern-.05em{\sc i\kern-.025em b}\kern-.08em
    T\kern-.1667em\lower.7ex\hbox{E}\kern-.125emX}}

\begin{document}

\title{PCDT: A Predictive Cognitive Digital Twin Framework for Intelligent and Autonomous 6G Network Ecosystems}

\author{\IEEEauthorblockN{John Sengendo}
\IEEEauthorblockA{C.N.I.T and \\
University of Trento, ITALY.\\
Email: john.sengendo@unitn.it}
\and
\IEEEauthorblockN{Andreas Kassler}
\IEEEauthorblockA{
Deggendorf Institute of Technology, Germany, and\\
Karlstad University, Sweden.\\
Email: andreas.kassler@th-deg.de}
\and
\IEEEauthorblockN{Fabrizio Granelli}
\IEEEauthorblockA{C.N.I.T and \\
University of Trento, ITALY.\\
Email: fabrizio.granelli@unitn.it}

}

\maketitle

\begin{abstract}
Future 6G networks are expected to operate as intelligent and autonomous ecosystems
where monitoring, prediction, and control are integrated into continuous
self-optimization loops. However, many digital-twin-based network management
approaches still act mainly as synchronized replicas of the network. They observe
the state, report degradation, and trigger corrective action only after performance
risk has appeared. This leaves a gap between the vision of a cognitive digital
twin (CDT) and the behavior of conventional reactive control loops. In this paper, we propose a Predictive Cognitive Digital Twin (PCDT) 
framework that closes the loop from observation to predictive cognition to
proactive resource control. PCDT maintains a persistent traffic world model, forecasts near-future load, and allocates capacity for the predicted horizon peak before a Service Level Agreement (SLA) violation occurs. Evaluated on a real-world traffic trace, PCDT achieves the lowest allocation error among all benchmarked baseline frameworks, reducing MAE by 43\% and RMSE by 32\% relative to the best baseline (reactive control). Relative to the threshold heuristic, the only baseline with zero violations, PCDT reduces mean allocated capacity by 50\%, reconfiguration churn by 56\%, and total operating cost by 50\%, indicating substantially more efficient performance. These results show the proposed framework advances the digital twin (DT) operation mechanism from a passive representation towards a cognitive, proactive control, and "intent-aware" mechanism aligned with autonomous 6G network vision.
\end{abstract}


\section{Introduction}

The transition towards 6G is not only a radio-access evolution, it is also an
operational transformation \cite{yang2024large, huang2025digital}. Future networks are expected to support massive
heterogeneity, strict service-level objectives, energy constraints, and dynamic
application demands. Managing such environments with static rules or
"human-in-the-loop" reactions is increasingly unrealistic. The emerging vision is an intelligent and autonomous network ecosystem, a system that senses its state, understands service intent, predicts near-future risk, and optimizes itself before user experience degrades \cite{bilen2025knowledge}.

Digital twins (DTs) are widely viewed as an enabling technology for this vision \cite{lin20236g, kuruvatti2022empowering}. Moreover as studies underscore \cite{Chen},  digital twins can not only virtually replicate a system’s detailed operations, but also analyze the current condition, predict the future behavior, and refine control optimization \cite{Chen}. A network digital twin (NDT) can synchronize telemetry, represent the current state of a network
cell or slice, and support safe evaluation of control decisions \cite{tariq2022toward}. 

A digital replica that
only mirrors the present is not sufficient for autonomous operation. If the
controller waits until a traffic surge has already created congestion, the twin
has helped identify the problem but has not prevented it. This limitation is the
core motivation for cognitive network digital twins. The twin must become an active cognitive entity that reasons about the future and acts in advance \cite{bilen2025knowledge}. As additionally noted by authors in \cite{okegbile}, prediction-enhanced frameworks can trigger virtual twin evolution in advance using expected states, reducing synchronization delays to near-zero \cite{okegbile}.

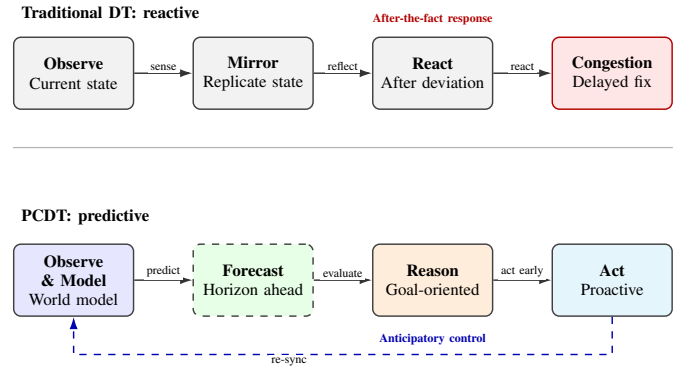
\begin{figure}[h]
\centering
\resizebox{\columnwidth}{!}{
\begin{tikzpicture}[
    node distance=0.55cm and 0.85cm,
    stage/.style={
        rectangle,
        rounded corners=3pt,
        draw=black!75,
        line width=0.6pt,
        minimum width=1.75cm,
        minimum height=1.05cm,
        text width=1.6cm,
        align=center,
        font=\scriptsize,
        inner sep=2pt
    },
    traditional/.style={stage, fill=gray!10},
    world/.style={stage, fill=blue!10},
    prediction/.style={stage, fill=green!10, dashed},
    reasoning/.style={stage, fill=orange!14},
    action/.style={stage, fill=cyan!9},
    adverse/.style={stage, fill=red!10, draw=red!70!black},
    flow/.style={
        -{Latex[length=1.8mm,width=1.2mm]},
        line width=0.6pt,
        draw=black!80
    },
    feedback/.style={flow, dashed, draw=blue!70!black},
    rowtitle/.style={
        font=\scriptsize\bfseries,
        anchor=west
    },
    arrowlabel/.style={
        midway, above, font=\tiny, fill=white, inner sep=1pt
    }
]
\node[traditional] (observe-t) {\textbf{Observe}\\Current state};
\node[traditional, right=of observe-t] (mirror) {\textbf{Mirror}\\Replicate state};
\node[traditional, right=of mirror] (reactive) {\textbf{React}\\After deviation};
\node[adverse, right=of reactive] (congestion) {\textbf{Congestion}\\Delayed fix};
\node[rowtitle, above=0.15cm of observe-t.north west, anchor=south west]
    {Traditional DT: reactive};
\draw[flow] (observe-t) -- node[arrowlabel] {sense} (mirror);
\draw[flow] (mirror) -- node[arrowlabel] {reflect} (reactive);
\draw[flow] (reactive) -- node[arrowlabel] {react} (congestion);
\node[above=0.05cm of reactive, font=\tiny\bfseries, text=red!70!black]
    {After-the-fact response};
\draw[black!30, line width=0.5pt]
    ($(observe-t.south west)+(0,-0.55)$) -- ($(congestion.south east)+(0,-0.55)$);
\node[world, below=1.95cm of observe-t] (world) {\textbf{Observe \& Model}\\World model};
\node[prediction, right=of world] (forecast) {\textbf{Forecast}\\Horizon ahead};
\node[reasoning, right=of forecast] (reason) {\textbf{Reason}\\Goal-oriented};
\node[action, right=of reason] (proactive) {\textbf{Act}\\Proactive};
\node[rowtitle, above=0.15cm of world.north west, anchor=south west]
    {PCDT: predictive};
\draw[flow] (world) -- node[arrowlabel] {predict} (forecast);
\draw[flow] (forecast) -- node[arrowlabel] {evaluate} (reason);
\draw[flow] (reason) -- node[arrowlabel] {act early} (proactive);
\node[below=0.12cm of reason, font=\tiny\bfseries, text=blue!70!black]
    {Anticipatory control};
\draw[feedback]
    (proactive.south) -- ++(0,-0.55)
    -| node[pos=0.30, below, font=\tiny, fill=white, inner sep=1pt] {re-sync}
    (world.south);
\end{tikzpicture}}
\caption{Traditional reactive DT versus the proposed PCDT. The PCDT forecasts network conditions and enables proactive capacity control before violation occurs.}
\label{fig:traditional_dt_vs_pcdt}
\end{figure}
The novelty of PCDT stems from the joint introduction of four elements, combined
into a single closed-loop cognitive architecture: (i) a \emph{persistent
world model} that maintains a continuously updated state of network traffic,
rather than a snapshot replicated on demand; (ii) \emph{uncertainty-aware
horizon planning}, in which the forecasting layer estimates a distribution
over horizon-ahead load rather than a single deterministic point estimate,
allowing the controller to reason over confidence rather than a single
predicted value; (iii) a \emph{proactive goal-oriented controller} that
maps this uncertainty-aware forecast directly onto a capacity decision
before deviation occurs, rather than triggering a corrective action after
a threshold is breached, and (iv) \emph{continuous re-synchronization},
in which the world model is updated using the realized outcome of each
action, closing the loop rather than treating prediction and actuation as
a one-shot process. Prior network digital twin frameworks and surveys (e.g., \cite{Verdecchia, Ruah}) have typically treated prediction, monitoring, and control as separate capabilities, whereas PCDT integrates them into a single closed loop
as depicted in Figure.~\ref{fig:traditional_dt_vs_pcdt}.

Additionally, our study focuses on a concrete control problem. At each time step, a network
cell or slice receives a load demand and the controller chooses allocated
capacity. Under-allocation creates service level agreement violations, over-allocation wastes
resources, and frequent changes create reconfiguration churn. A reactive digital
twin can track the previous load, but it naturally lags surges. Static
overprovisioning can reduce risk, but it wastes capacity \cite{lusani2025framework}. Handcrafted threshold
logic can anticipate some ramps, but often increases churn. Our proposed mechanism is designed
to occupy the middle ground, proactive enough to avoid violations, but efficient
and stable enough to be useful for autonomous network operation \cite{bilen2025knowledge}. Below we underscore the key contributions of this paper:
\begin{itemize}
\item We formulate a predictive cognitive digital twin architecture integrating
persistent world modeling, uncertainty-aware forecasting, and proactive
capacity allocation.
\item We derive a multi-objective control formulation jointly minimizing
provisioning cost, SLA violations, and reconfiguration churn.
\item We evaluate on real cellular traffic traces and demonstrate 43\% lower MAE,
50\% lower operating cost compared to
baseline methods.
\item Additionally, our proposed framework and results demonstrate how PCDT aligns with 6G and 3GPP properties and roadmap.
\end{itemize}

The rest of the paper is organized as follows: Section II reviews related work, Section III presents the experimentation framework and methodology, Section IV discusses the results, and Section V concludes the paper.

\section{Related Work}

\subsection{Network Digital Twins}

Digital twins emerged as virtual replicas of physical systems and have
since become a key enabler technology in various sectors such as manufacturing, automotive, healthcare, and mobile networks \cite{mihai2022digital}. In the mobile network eco-system, digital twins can model network topologies, mobile traffic, radio resources, service state, and operational policies, which makes them useful for monitoring and "what-if" case scenario analysis. Works in \cite{Verdecchia} review primary studies and show that most network digital twin frameworks still focus on network performance optimization, with fewer studies addressing deeper closed-loop autonomy \cite{Verdecchia}. An in-depth review in \cite{wang} examines Internet of DTs (IoDT) architecture and identifies data synchronization between physical and digital spaces as a core requirement and challenge, with service requirements ranging from real-time (millisecond) to delay-tolerant (minute) timescales \cite{wang}. Additionally, authors in \cite{Gonzalez} propose a NDT architecture that maps network-management information into an interoperable model for control and management, yet the emphasis remains on lifecycle mirroring and closed-loop automation support. With in the same lines, works in \cite{Hanzhi} explicitly study synchronization between the physical network and the digital twin, with their framework still handling synchronization, prediction, and resource allocation separately. This leaves a gap for novel frameworks such as PCDT, that not only synchronize, but also are uncertainty-aware, proactive, and continuously update through action outcomes.

\subsection{Traffic Forecasting and Control}

Traffic forecasting has been studied using persistence baselines frameworks, autoregressive
models, neural networks, recurrent models, temporal convolutional networks,
attention-based models, and graph temporal architectures. As authors in \cite{nhu2022dynamic} demonstrate, attention-based models can achieve high prediction accuracy for multivariate time series, while works in \cite{patidar2025spatio} show that graph neural networks with optimized attention mechanisms effectively capture spatial-temporal dependencies in mobile networks. Moreover, spatiotemporal network traffic captures both spatial and temporal traffic dynamics, enabling a deeper understanding of network behavior and prediction \cite{qiu2023network}. With trained models improving prediction accuracy and forecasting, this can enable proactive network management \cite{ferreira2021forecasting}, but forecasting accuracy alone does not guarantee
good network control. PCDT is positioned novelly, because it connects uncertainty-aware forecasting directly to a capacity decision and then feeds the realized outcome back into the twin.

Additionally, resource control state-of-the-art includes frameworks such as; static provisioning, threshold-based
heuristics, model predictive control, reinforcement learning, and intent-based
automation. Studies presented by the authors in \cite{wu2021deep}, show that reinforcement learning approaches such as Actor-Critic methods can be combined with traffic forecasting for effective base station sleep control. Static frameworks are stable but inefficient, reactive methods are
efficient in calm periods but vulnerable to rapid changes, learned policies can
be powerful but require more training and validation as shown in \cite{ferreira2021forecasting}. PCDT is positioned as a compact, interpretable framework that
can later absorb stronger forecasters or policy learners while preserving the
observe-predict-act work flow.

\subsection{Autonomous 6G Network Ecosystems}
ETSI ZSM, O-RAN, and AI-native network studies all point towards closed-loop automation in 6G. In practice however, much of the literature still focuses on enabling the plumbing of autonomy rather than integrating cognition, prediction, and control into one loop. OpenTwin presented in \cite{hossen2026opentwin}, for example, uses a deviation-aware scoring mechanism that automatically triggers re-synchronization when network drift is detected, which is a clear step towards self-maintaining twins \cite{hossen2026opentwin}. The same pattern appears in O-RAN-oriented works in \cite{brik2024explainable}, where authors proposes an open-source NDT framework for O-RAN and emphasize real-time monitoring, simulation, and optimization. These works show that the eco-system is already moving towards autonomy, but they also reveal the remaining gap where existing mechanisms often manage observability, synchronization, or resource allocation separately. PCDT differs by providing a novel closed-loop cognitive mechanism. In the next section, we present our experimentation framework.

\begin{figure}
    \centering
    \input{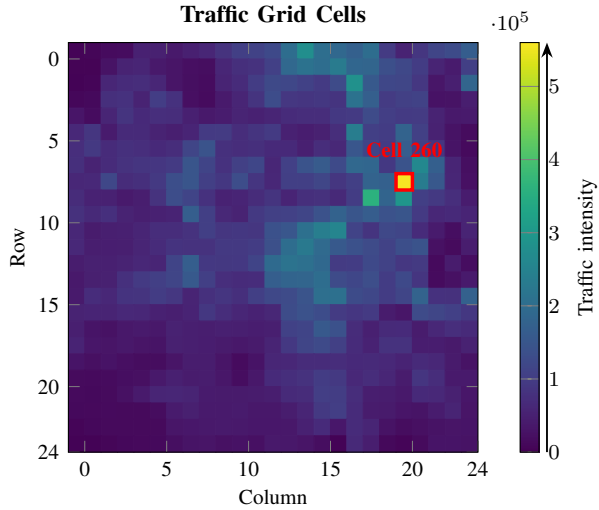}
    \caption{Traffic grid cells highlighting the high traffic cell 260.}
    \label{fig:traffic_grids}
\end{figure}

\begin{figure}
  \centering
  \resizebox{\columnwidth}{!}{
  \begin{tikzpicture}
  \begin{axis}[
    width=15cm, height=8cm,
    ylabel={Normalized load},
    xlabel={Time},
    xlabel style={font=\large},
    ylabel style={font=\large},
    tick label style={font=\large},
    xmin=0, xmax=351,
    ymin=0, ymax=1.02,
    xtick={0,72,144,216,288},
    xticklabels={
      {01-Nov\\00:00},{01-Nov\\12:00},
      {02-Nov\\00:00},{02-Nov\\12:00},
      {03-Nov\\00:00}
    },
    xticklabel style={font=\large, align=center},
    grid=major,
    grid style={gray!20},
    legend style={
      at={(1.01,1)},
      anchor=north west,
      font=\large,
      draw=gray!50,
      fill=white,
      row sep=2pt,
    },
    legend cell align=left,
  ]
    \addplot[fill=green!15, draw=none, area legend] coordinates {
      (0,0.0346)(1,0.0335)(2,0.0409)(3,0.0309)(4,0.0206)(5,0.0279)(6,0.0135)(7,0.0160)(8,0.0104)(9,0.0163)(10,0.0088)(11,0.0073)(12,0.0053)(13,0.0070)(14,0.0033)(15,0.0045)(16,0.0085)(17,0.0077)(18,0.0027)(19,0.0041)(20,0.0070)(21,0.0030)(22,0.0026)(23,0.0039)(24,0.0015)(25,0.0008)(26,0.0033)(27,0.0000)(28,0.0024)(29,0.0036)(30,0.0013)(31,0.0050)(32,0.0081)(33,0.0003)(34,0.0022)(35,0.0038)(36,0.0087)(37,0.0054)(38,0.0081)(39,0.0077)(40,0.0103)(41,0.0121)(42,0.0079)(43,0.0080)(44,0.0121)(45,0.0079)(46,0.0168)(47,0.0182)(48,0.0137)(49,0.0108)(50,0.0105)(51,0.0171)(52,0.0221)(53,0.0171)(54,0.0147)(55,0.0303)(56,0.0338)(57,0.0270)(58,0.0363)(59,0.0672)(60,0.0559)(61,0.0707)(62,0.0747)(63,0.1435)(64,0.1214)(65,0.1172)(66,0.1806)(67,0.1529)(68,0.1994)(69,0.2582)(70,0.2179)(71,0.2418)(72,0.2504)(73,0.2768)(74,0.3328)(75,0.3511)(76,0.3233)(77,0.4105)(78,0.3437)(79,0.3834)(80,0.4042)(81,0.4014)(82,0.3961)(83,0.4300)(84,0.4354)(85,0.4304)(86,0.4064)(87,0.4494)(88,0.4310)(89,0.4361)(90,0.5897)(91,0.4720)(92,0.4344)(93,0.4887)(94,0.4423)(95,0.5459)(96,0.5392)(97,0.6182)(98,0.5537)(99,0.5495)(100,0.5708)(101,0.5664)(102,0.5526)(103,0.5502)(104,0.5862)(105,0.5581)(106,0.6246)(107,0.5956)(108,0.6024)(109,0.5615)(110,0.4796)(111,0.4794)(112,0.5398)(113,0.4463)(114,0.4182)(115,0.4225)(116,0.3728)(117,0.3702)(118,0.3478)(119,0.3330)(120,0.2993)(121,0.2750)(122,0.2222)(123,0.2016)(124,0.2101)(125,0.2158)(126,0.1703)(127,0.1425)(128,0.1355)(129,0.1383)(130,0.1041)(131,0.1172)(132,0.1266)(133,0.1343)(134,0.1330)(135,0.1056)(136,0.0743)(137,0.0720)(138,0.0570)(139,0.0692)(140,0.0714)(141,0.0802)(142,0.0613)(143,0.0473)(144,0.0247)(145,0.0551)(146,0.0444)(147,0.0209)(148,0.0264)(149,0.0161)(150,0.0093)(151,0.0104)(152,0.0141)(153,0.0111)(154,0.0118)(155,0.0076)(156,0.0046)(157,0.0099)(158,0.0072)(159,0.0036)(160,0.0066)(161,0.0052)(162,0.0039)(163,0.0056)(164,0.0043)(165,0.0029)(166,0.0054)(167,0.0049)(168,0.0021)(169,0.0119)(170,0.0102)(171,0.0085)(172,0.0231)(173,0.0139)(174,0.0186)(175,0.0100)(176,0.0099)(177,0.0057)(178,0.0052)(179,0.0049)(180,0.0016)(181,0.0055)(182,0.0063)(183,0.0078)(184,0.0084)(185,0.0107)(186,0.0109)(187,0.0093)(188,0.0077)(189,0.0051)(190,0.0099)(191,0.0100)(192,0.0087)(193,0.0092)(194,0.0107)(195,0.0220)(196,0.0167)(197,0.0200)(198,0.0283)(199,0.0367)(200,0.0941)(201,0.0422)(202,0.0774)(203,0.0690)(204,0.0853)(205,0.0962)(206,0.1277)(207,0.1422)(208,0.2674)(209,0.2360)(210,0.2353)(211,0.2896)(212,0.2659)(213,0.3099)(214,0.3992)(215,0.4671)(216,0.4255)(217,0.4020)(218,0.4665)(219,0.4496)(220,0.5786)(221,0.5903)(222,0.4693)(223,0.5114)(224,0.5722)(225,0.6632)(226,0.5857)(227,1.0000)(228,0.5642)(229,0.5596)(230,0.7009)(231,0.6153)(232,0.6123)(233,0.5681)(234,0.6651)(235,0.6618)(236,0.6673)(237,0.6817)(238,0.5881)(239,0.6598)(240,0.6372)(241,0.6596)(242,0.7128)(243,0.6787)(244,0.6593)(245,0.6584)(246,0.6388)(247,0.6136)(248,0.6028)(249,0.6670)(250,0.5720)(251,0.6366)(252,0.6891)(253,0.5123)(254,0.5669)(255,0.5931)(256,0.4913)(257,0.4431)(258,0.4659)(259,0.4226)(260,0.4534)(261,0.4411)(262,0.3669)(263,0.3960)(264,0.3311)(265,0.3272)(266,0.2523)(267,0.2265)(268,0.2371)(269,0.1976)(270,0.2320)(271,0.1366)(272,0.1906)(273,0.1919)(274,0.1246)(275,0.1817)(276,0.1634)(277,0.1604)(278,0.1399)(279,0.1344)(280,0.1875)(281,0.0965)(282,0.0936)(283,0.0951)(284,0.0741)(285,0.0831)(286,0.0717)(287,0.0600)(288,0.0403)(289,0.0523)(290,0.0726)(291,0.0618)(292,0.0364)(293,0.0475)(294,0.0758)(295,0.0502)(296,0.0229)(297,0.0178)(298,0.0179)(299,0.0099)(300,0.0180)(301,0.0179)(302,0.0630)(303,0.0092)(304,0.0101)(305,0.0044)(306,0.0119)(307,0.0160)(308,0.0141)(309,0.0084)(310,0.0088)(311,0.0036)(312,0.0074)(313,0.0116)(314,0.0052)(315,0.0099)(316,0.0056)(317,0.0029)(318,0.0049)(319,0.0068)(320,0.0017)(321,0.0067)(322,0.0092)(323,0.0024)(324,0.0092)(325,0.0077)(326,0.0030)(327,0.0110)(328,0.0085)(329,0.0060)(330,0.0109)(331,0.0114)(332,0.0070)(333,0.0185)(334,0.0132)(335,0.0072)(336,0.0136)(337,0.0160)(338,0.0095)(339,0.0136)(340,0.0162)(341,0.0464)(342,0.0218)(343,0.0206)(344,0.0195)(345,0.0300)(346,0.0766)(347,0.0398)(348,0.0449)(349,0.0536)(350,0.1078)(351,0.0747)
    } \closedcycle;
    \addlegendentry{Traffic variation};
    \addplot[blue, line width=1pt] coordinates {
      (0,0.0346)(1,0.0335)(2,0.0409)(3,0.0309)(4,0.0206)(5,0.0279)(6,0.0135)(7,0.0160)(8,0.0104)(9,0.0163)(10,0.0088)(11,0.0073)(12,0.0053)(13,0.0070)(14,0.0033)(15,0.0045)(16,0.0085)(17,0.0077)(18,0.0027)(19,0.0041)(20,0.0070)(21,0.0030)(22,0.0026)(23,0.0039)(24,0.0015)(25,0.0008)(26,0.0033)(27,0.0000)(28,0.0024)(29,0.0036)(30,0.0013)(31,0.0050)(32,0.0081)(33,0.0003)(34,0.0022)(35,0.0038)(36,0.0087)(37,0.0054)(38,0.0081)(39,0.0077)(40,0.0103)(41,0.0121)(42,0.0079)(43,0.0080)(44,0.0121)(45,0.0079)(46,0.0168)(47,0.0182)(48,0.0137)(49,0.0108)(50,0.0105)(51,0.0171)(52,0.0221)(53,0.0171)(54,0.0147)(55,0.0303)(56,0.0338)(57,0.0270)(58,0.0363)(59,0.0672)(60,0.0559)(61,0.0707)(62,0.0747)(63,0.1435)(64,0.1214)(65,0.1172)(66,0.1806)(67,0.1529)(68,0.1994)(69,0.2582)(70,0.2179)(71,0.2418)(72,0.2504)(73,0.2768)(74,0.3328)(75,0.3511)(76,0.3233)(77,0.4105)(78,0.3437)(79,0.3834)(80,0.4042)(81,0.4014)(82,0.3961)(83,0.4300)(84,0.4354)(85,0.4304)(86,0.4064)(87,0.4494)(88,0.4310)(89,0.4361)(90,0.5897)(91,0.4720)(92,0.4344)(93,0.4887)(94,0.4423)(95,0.5459)(96,0.5392)(97,0.6182)(98,0.5537)(99,0.5495)(100,0.5708)(101,0.5664)(102,0.5526)(103,0.5502)(104,0.5862)(105,0.5581)(106,0.6246)(107,0.5956)(108,0.6024)(109,0.5615)(110,0.4796)(111,0.4794)(112,0.5398)(113,0.4463)(114,0.4182)(115,0.4225)(116,0.3728)(117,0.3702)(118,0.3478)(119,0.3330)(120,0.2993)(121,0.2750)(122,0.2222)(123,0.2016)(124,0.2101)(125,0.2158)(126,0.1703)(127,0.1425)(128,0.1355)(129,0.1383)(130,0.1041)(131,0.1172)(132,0.1266)(133,0.1343)(134,0.1330)(135,0.1056)(136,0.0743)(137,0.0720)(138,0.0570)(139,0.0692)(140,0.0714)(141,0.0802)(142,0.0613)(143,0.0473)(144,0.0247)(145,0.0551)(146,0.0444)(147,0.0209)(148,0.0264)(149,0.0161)(150,0.0093)(151,0.0104)(152,0.0141)(153,0.0111)(154,0.0118)(155,0.0076)(156,0.0046)(157,0.0099)(158,0.0072)(159,0.0036)(160,0.0066)(161,0.0052)(162,0.0039)(163,0.0056)(164,0.0043)(165,0.0029)(166,0.0054)(167,0.0049)(168,0.0021)(169,0.0119)(170,0.0102)(171,0.0085)(172,0.0231)(173,0.0139)(174,0.0186)(175,0.0100)(176,0.0099)(177,0.0057)(178,0.0052)(179,0.0049)(180,0.0016)(181,0.0055)(182,0.0063)(183,0.0078)(184,0.0084)(185,0.0107)(186,0.0109)(187,0.0093)(188,0.0077)(189,0.0051)(190,0.0099)(191,0.0100)(192,0.0087)(193,0.0092)(194,0.0107)(195,0.0220)(196,0.0167)(197,0.0200)(198,0.0283)(199,0.0367)(200,0.0941)(201,0.0422)(202,0.0774)(203,0.0690)(204,0.0853)(205,0.0962)(206,0.1277)(207,0.1422)(208,0.2674)(209,0.2360)(210,0.2353)(211,0.2896)(212,0.2659)(213,0.3099)(214,0.3992)(215,0.4671)(216,0.4255)(217,0.4020)(218,0.4665)(219,0.4496)(220,0.5786)(221,0.5903)(222,0.4693)(223,0.5114)(224,0.5722)(225,0.6632)(226,0.5857)(227,1.0000)(228,0.5642)(229,0.5596)(230,0.7009)(231,0.6153)(232,0.6123)(233,0.5681)(234,0.6651)(235,0.6618)(236,0.6673)(237,0.6817)(238,0.5881)(239,0.6598)(240,0.6372)(241,0.6596)(242,0.7128)(243,0.6787)(244,0.6593)(245,0.6584)(246,0.6388)(247,0.6136)(248,0.6028)(249,0.6670)(250,0.5720)(251,0.6366)(252,0.6891)(253,0.5123)(254,0.5669)(255,0.5931)(256,0.4913)(257,0.4431)(258,0.4659)(259,0.4226)(260,0.4534)(261,0.4411)(262,0.3669)(263,0.3960)(264,0.3311)(265,0.3272)(266,0.2523)(267,0.2265)(268,0.2371)(269,0.1976)(270,0.2320)(271,0.1366)(272,0.1906)(273,0.1919)(274,0.1246)(275,0.1817)(276,0.1634)(277,0.1604)(278,0.1399)(279,0.1344)(280,0.1875)(281,0.0965)(282,0.0936)(283,0.0951)(284,0.0741)(285,0.0831)(286,0.0717)(287,0.0600)(288,0.0403)(289,0.0523)(290,0.0726)(291,0.0618)(292,0.0364)(293,0.0475)(294,0.0758)(295,0.0502)(296,0.0229)(297,0.0178)(298,0.0179)(299,0.0099)(300,0.0180)(301,0.0179)(302,0.0630)(303,0.0092)(304,0.0101)(305,0.0044)(306,0.0119)(307,0.0160)(308,0.0141)(309,0.0084)(310,0.0088)(311,0.0036)(312,0.0074)(313,0.0116)(314,0.0052)(315,0.0099)(316,0.0056)(317,0.0029)(318,0.0049)(319,0.0068)(320,0.0017)(321,0.0067)(322,0.0092)(323,0.0024)(324,0.0092)(325,0.0077)(326,0.0030)(327,0.0110)(328,0.0085)(329,0.0060)(330,0.0109)(331,0.0114)(332,0.0070)(333,0.0185)(334,0.0132)(335,0.0072)(336,0.0136)(337,0.0160)(338,0.0095)(339,0.0136)(340,0.0162)(341,0.0464)(342,0.0218)(343,0.0206)(344,0.0195)(345,0.0300)(346,0.0766)(347,0.0398)(348,0.0449)(349,0.0536)(350,0.1078)(351,0.0747)
    };
    \addlegendentry{Normalized load};
  \end{axis}
\end{tikzpicture}}
  \caption {Normalized traffic load trace for cell 260. Green shading denoting load variation.}
  \label{fig:traffic_load}
\end{figure}
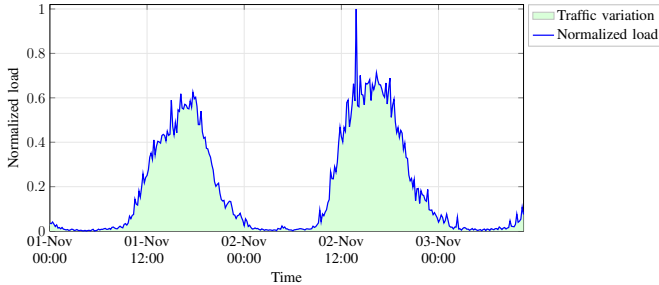

\section{Experimental Framework}

\subsection{Dataset and Preprocessing}

As earlier highlighted, our framework uses a real-world dataset capturing cellular traffic activities \cite{barlacchi2015multi} aggregated at intervals of 10 minutes  \cite{barlacchi2015multi}. The metrics reflected within the dataset particularly traffic intensity are critical for network capacity planning \cite{kitahara2015practical}, as they directly reflect user demand and determine service-level requirements \cite{Hampel}. The used network traffic additionally exhibits spatiotemporal variability, whose characterization is fundamental for the design, optimization, and efficient resource management of mobile cellular networks, as it enables accurate modeling of user demand across both spatial and temporal dimensions \cite{Bothe}. Additionally, the dataset has been applied in traffic forecasting studies such as in \cite{11223325}.
Moreover, the dataset's structure pairing a \texttt{time} field with a \texttt{cell\_id} makes it well suited for isolating and studying the behavior of any single cell within the network and the effect of neighboring cells. For illustration of the grid traffic, we extracted a $25\times25$ grid region from the entire dataset as shown in Figure~\ref{fig:traffic_grids}, which captures the spatial distribution of traffic intensity across the area, showing different cells depicting varying traffic intensities.

As illustrated in the grid, cell 260 emerged as the cell with the highest traffic intensity, its total load exceeding that of the next-busiest cells, marking it as a high-demand urban hotspot where capacity management is most highly stressed. Beyond its raw volume, the cell's load trace is markedly non-stationary with intensity varying sharply across the day, escalating through sustained peak windows as shown in Figure.~\ref{fig:traffic_load}, which also traces the cell's normalized load over time, with green shading marking the envelope of traffic variation. These properties make Cell 260 a suitable testbed for contrasting proactive and reactive allocation strategies. 

During pre-processing, traffic intensity was normalized to the $[0,1]$ range using a min-max scaling approach a standard preprocessing step that brings variables onto a common scale \cite{panday2022selective}, as shown in Eq.~\eqref{eq:minmax}. Additionally, the min-max scaling approach was selected to place variables on a common scale for fair comparison, a common goal of normalization in time-series analysis \cite{lima2023large}.

\begin{equation}
x_t = \frac{y_t - \min(y)}{\max(y) - \min(y) + \epsilon},
\label{eq:minmax}
\end{equation}

where $\epsilon = 10^{-9}$ is a smoothing term added to the denominator to prevent division by zero in cases where a cell exhibits no variance over the observation window (i.e., $\max(y) = \min(y)$).

\begin{table}[t]
\caption{Experimental Configurations}
\label{tab:params}
\centering
\begin{tabular}{@{}ll@{}}
\toprule
\textbf{Component} & \textbf{Parameters} \\
\midrule
Dataset &
\makecell[l]{Cell 260 selected (busiest)\\
70\% / 15\% / 15\% train/val/test} \\
\addlinespace
Forecasting &
\makecell[l]{Lookback $L{=}12$, horizon $H{=}6$\\
Persistence, Ridge ($\alpha{=}1.0$), MLP (128-64-32)\\
Selected by minimum validation RMSE} \\
\addlinespace
Cost function &
\makecell[l]{$\alpha{=}1.0$ (capacity), $\beta{=}8.0$ (violation),\\
$\gamma{=}0.3$ (reconfiguration)} \\
\addlinespace
PCDT (proposed) &
\makecell[l]{Percentile $p{=}95$, base margin $\delta_0{=}0.02$\\
Margin sensitivity $\lambda{=}0.01$, smoothing $\alpha{=}0.6$} \\
\addlinespace
Baselines &
\makecell[l]{Static: 98th pct.\quad Reactive: margin 0.05\\
Threshold: margin 0.08, slope gain 3.0} \\
\bottomrule
\end{tabular}
\end{table}

\subsection{Predictive Cognition Layer}

At time $t$, PCDT builds a feature vector from the previous load $L$ values and simple temporal context:
\begin{equation}
\phi_t =
[x_{t-L},\ldots,x_{t-1},\,s_t,\,\mu_t,\,m_t,\,\sigma_t,\,\sin\theta_t,\,\cos\theta_t,\,h_t],
\label{eq:features}
\end{equation}
where $s_t$ is the \emph{window slope}, a first-order finite-difference
estimate of the local trend computed as $s_t=(x_{t-1}-x_{t-L})/(L-1)$,
indicating whether load is rising or falling over the lookback horizon,
$\mu_t$ and $m_t$ are the window \emph{mean} and \emph{maximum}, standard rolling-window summary statistics that capture the recent typical and peak
load levels, following common time-series feature-engineering techniques. $\sigma_t$ is the window \emph{volatility} (standard deviation), measuring the local variability in the recent load history and providing the forecaster with a signal for regime instability; a related quantity, the standard deviation of the forecaster's own predicted
horizon ($\hat{\sigma}_t$), is what directly drives the adaptive safety margin. 

$\theta_t$ is the \emph{daily phase angle}, $\theta_t=2\pi(t \bmod P)/P$ for a period
$P{=}144$ ten-minute steps (24~h), whose encoding as
$(\sin\theta_t,\cos\theta_t)$ maps cyclical time-of-day onto a continuous, discontinuity-free representation, a technique shown to be effective for
diurnal cellular traffic specifically; and $h_t\in[0,1)$ is the complementary \emph{linear hour-of-day phase},
$h_t=(t \bmod P)/P$, a monotonic (non-cyclical) time indicator that lets
the model separately learn absolute time-of-day effects (e.g., midday
vs.\ midnight baselines) alongside the cyclical encoding above. The same technique is applied identically at training
and inference time to avoid train/serve skew. The forecaster outputs a
horizon of future load estimates:
\begin{equation}
\hat{\mathbf{x}}_{t+1:t+H}=f_{\omega}(\phi_t).
\label{eq:forecast}
\end{equation}

Three candidate forecasters were trained on the identical feature
representation: a persistence baseline, a Ridge-regularized linear
autoregressor ($\alpha{=}1.0$), and a 3-layer MLP (128--64--32 units, ReLU,
early stopping). Model selection was performed by minimum RMSE on the
validation split, and the selected model evaluated once, out-of-sample,
on the held-out test split. Across all runs of this pipeline, the linear
autoregressive model was consistently selected.

\subsection{Goal-Oriented Control Layer}
The controller chooses a capacity $c_t$. If the realized load exceeds capacity,
the system incurs an SLA violation. If capacity exceeds load, resources are
wasted. If capacity changes too often, the system incurs reconfiguration cost. The total objective is:
\begin{equation}
J = \sum_{t\in\mathcal{T}} \alpha c_t
  + \beta [x_t-c_t]^+
  + \gamma |c_t-c_{t-1}|,
\label{eq:cost}
\end{equation}
where $[z]^+=\max(z,0)$, $\alpha=1.0$, $\beta=8.0$, and $\gamma=0.3$ as indicated in Table~\ref{tab:params}.
PCDT converts the forecast into capacity through three coupled mechanisms,
shown in Algorithm~\ref{alg:pcdt}. First, it uses the 95th percentile of the
horizon predictions rather than the maximum, trading a small tail-risk for
lower over-provisioning. Second, it scales the safety margin with forecast
uncertainty: a higher standard deviation of $\hat{\mathbf{x}}$ widens the
buffer automatically. 
Third, exponential smoothing damps rapid
reconfigurations without sacrificing responsiveness:
\begin{equation}
\tilde{c}_t =
\max\!\left(\hat{x}_{t,p_{95}},\, x_{t-1}\right) + \delta_t,
\qquad
c_t = \alpha\,\tilde{c}_t + (1-\alpha)\,c_{t-1},
\label{eq:pcdt}
\end{equation}
where $\hat{x}_{t,p_{95}}$ is the 95th percentile of the horizon $H$,
$\delta_t = \delta_0 + \lambda\,\sigma_{\hat{x}}$ is the adaptive margin
($\delta_0{=}0.02$, $\lambda{=}0.01$), and $\alpha{=}0.6$ is the smoothing
factor. Unlike $\max(\hat{x}_t)$, $\tilde{c}_t$ damps single-step forecast
noise, avoiding over-provisioning with no loss in reliability (max costs
1.3\% more capacity for the same violation rate). The PCDT does not wait
for congestion, it uses predicted future load to act before the next
violation.

\begin{algorithm}
\caption{PCDT Capacity Allocation at Step $t$}
\label{alg:pcdt}
\begin{algorithmic}[1]
\Require load history $x_{0:t}$, trained forecaster $f_\omega$,
         lookback $L$, horizon $H$, base margin $\delta_0$,
         smoothing $\alpha$, percentile $p$
\Ensure allocated capacity $c_t$
\If{$t < L$} \Comment{warm-up}
  \State $c_t \leftarrow x_t + \tfrac{\delta_0}{2}$
\Else
 \State $\phi_t \leftarrow \textsc{Featurize}(x_{t-L:t},\, t)$
  \State $\hat{\mathbf{x}} \leftarrow f_\omega(\phi_t)$
        \Comment{horizon predictions $\in\mathbb{R}^H$}
  \State $\hat{x}_{p} \leftarrow \text{percentile}(\hat{\mathbf{x}},\, p)$
  \State $\sigma_{\hat{x}} \leftarrow \text{std}(\hat{\mathbf{x}})$
  \State $\delta_t \leftarrow \delta_0 + 0.01\cdot\max(\sigma_{\hat{x}}, 0.01)$
        \Comment{adaptive margin}
  \State $\tilde{c}_t \leftarrow \max(\hat{x}_{p},\; x_{t-1}) + \delta_t$
  \State $c_t \leftarrow \alpha\,\tilde{c}_t + (1-\alpha)\,c_{t-1}$
        \Comment{exponential smoothing}
\EndIf
\State \Return $\max(c_t,\; 0.02)$
\end{algorithmic}
\end{algorithm}

\subsection{Benchmark Controllers}
PCDT was bechmarked against three frameworks: \begin{itemize}
    \item \textbf{Static overprovisioning}: allocates the 98th percentile test load, reflecting conservative capacity planning under traffic fluctuation \cite{Hampel}.
    \item \textbf{Reactive digital-twin control}: allocates the previous observed load plus a margin, reflecting a reactive mirroring strategy for traffic variation \cite{Hampel}.
    \item \textbf{Threshold heuristic}: increases capacity using a recent positive slope, consistent with threshold-based investment rules used in practical traffic-load planning \cite{kitahara2015practical}.
\end{itemize}
These baselines represent common operational alternatives: conservative provisioning, reactive mirroring, and handcrafted anticipation \cite{Hampel, kitahara2015practical}. In the following section, we discuss our results.

\begin{figure}[htbp]
  \centering
  \begin{tikzpicture}
    \begin{axis}[
      width=8.5cm, height=5.5cm,
      ylabel={Error value},
      ylabel style={font=\small},
      symbolic x coords={static, reactive, threshold, adapt-cdt},
      xtick=data,
      xticklabels={static, reactive, threshold, PCDT},
      xticklabel style={rotate=30, anchor=east, font=\small},
      yticklabel style={font=\small},
      tick label style={font=\small},
      ybar,
      bar width=14pt,
      enlarge x limits=0.25,
      ymin=0,
      ymax=0.11,
      grid=major,
      grid style={line width=0.3pt, draw=gray!30},
      legend style={
        at={(0.98,0.98)},
        anchor=north east,
        font=\small,
        draw=gray!40,
        fill=white,
      },
    ]

      \addplot[fill=blue!80, draw=black, line width=0.4pt] coordinates {
        (static,    0.0591)
        (reactive,  0.0489)
        (threshold, 0.0862)
        (adapt-cdt, 0.0277)
      };
      \addlegendentry{MAE}

      \addplot[fill=orange!70, draw=black, line width=0.4pt] coordinates {
        (static,    0.0617)
        (reactive,  0.0516)
        (threshold, 0.0874)
        (adapt-cdt, 0.0350)
      };
      \addlegendentry{RMSE}

    \end{axis}
  \end{tikzpicture}
  \caption{MAE and RMSE comparison across all the compared frameworks.}
  \label{fig:mae-rmse-bar}
\end{figure}
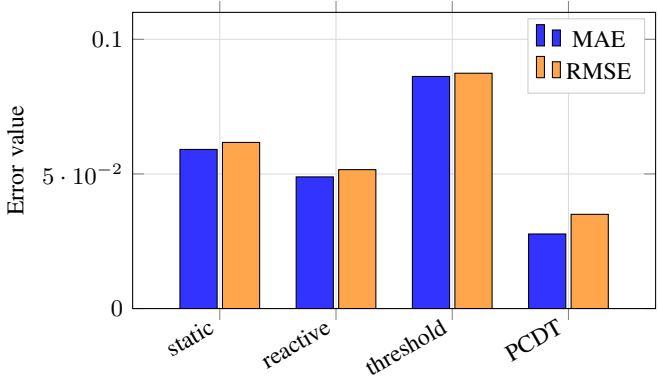

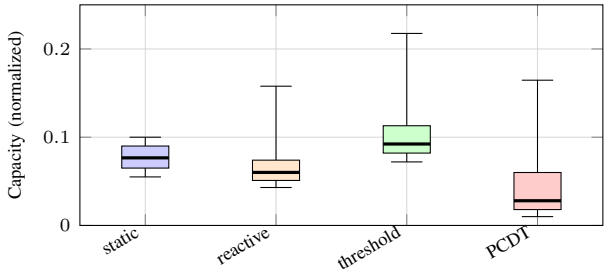
\begin{figure}[htbp]
  \centering
  \begin{tikzpicture}
    \begin{axis}[
      width=8.5cm, height=4.5cm,
      xmin=0.5, xmax=4.5,
      ymin=0.00, ymax=0.25,
      ylabel={Capacity (normalized)},
      ylabel style={font=\scriptsize},
      xlabel style={font=\scriptsize},
      yticklabel style={font=\scriptsize},
      xtick={1,2,3,4},
      xticklabels={static, reactive, threshold, PCDT},
      xticklabel style={rotate=30, anchor=east, font=\scriptsize},
      grid=major, grid style={gray!30},
      tick style={font=\scriptsize},
    ]

    \draw[fill=blue!20, draw=black] (0.82, 0.0650) rectangle (1.18, 0.0900);
    \draw[black, line width=1.2pt] (0.82, 0.0766) -- (1.18, 0.0766);
    \draw[black, thin] (1, 0.0550) -- (1, 0.0650);
    \draw[black, thin] (1, 0.0900) -- (1, 0.1000);
    \draw[black, thin] (0.88, 0.0550) -- (1.12, 0.0550);
    \draw[black, thin] (0.88, 0.1000) -- (1.12, 0.1000);

    \draw[fill=orange!20, draw=black] (1.82, 0.0510) rectangle (2.18, 0.0740);
    \draw[black, line width=1.2pt] (1.82, 0.0601) -- (2.18, 0.0601);
    \draw[black, thin] (2, 0.0430) -- (2, 0.0510);
    \draw[black, thin] (2, 0.0740) -- (2, 0.1578);
    \draw[black, thin] (1.88, 0.0430) -- (2.12, 0.0430);
    \draw[black, thin] (1.88, 0.1578) -- (2.12, 0.1578);

    \draw[fill=green!20, draw=black] (2.82, 0.0820) rectangle (3.18, 0.1130);
    \draw[black, line width=1.2pt] (2.82, 0.0923) -- (3.18, 0.0923);
    \draw[black, thin] (3, 0.0720) -- (3, 0.0820);
    \draw[black, thin] (3, 0.1130) -- (3, 0.2176);
    \draw[black, thin] (2.88, 0.0720) -- (3.12, 0.0720);
    \draw[black, thin] (2.88, 0.2176) -- (3.12, 0.2176);

    \draw[fill=red!20, draw=black] (3.82, 0.0180) rectangle (4.18, 0.0600);
    \draw[black, line width=1.2pt] (3.82, 0.0280) -- (4.18, 0.0280);
    \draw[black, thin] (4, 0.0100) -- (4, 0.0180);
    \draw[black, thin] (4, 0.0600) -- (4, 0.1646);
    \draw[black, thin] (3.88, 0.0100) -- (4.12, 0.0100);
    \draw[black, thin] (3.88, 0.1646) -- (4.12, 0.1646);

    \end{axis}
  \end{tikzpicture}
  \caption{Distribution of normalized capacity allocation across the frameworks.}
  \label{fig:capacity-boxplot}
\end{figure}

\begin{figure}[htbp]
  \centering
  \begin{tikzpicture}
    \begin{axis}[
      width=0.82\columnwidth, height=0.62\columnwidth,
      axis on top,
      enlargelimits=false,
      xtick={0,1,2,3,4},
      xticklabels={violations, capacity, severity, churn, cost},
      xticklabel style={rotate=45, anchor=north east, font=\scriptsize},
      ytick={0,1,2,3},
      yticklabels={static, reactive, threshold, PCDT},
      y dir=reverse,
      tick label style={font=\scriptsize},
      colormap={mygreen}{rgb255(0cm)=(20,90,60); rgb255(1cm)=(245,252,248)},
      point meta min=1, point meta max=4,
      colorbar,
      colorbar style={ylabel={Rank}, font=\scriptsize, width=0.2cm},
      nodes near coords={\pgfmathprintnumber[fixed, precision=0]{\pgfplotspointmeta}},
      every node near coord/.append style={font=\scriptsize, text=black},
    ]
      \addplot[matrix plot, mesh/cols=5, point meta=explicit] table[meta=C] {
        x y C
        0 0 4
        1 0 3
        2 0 4
        3 0 1
        4 0 3
        0 1 3
        1 1 2
        2 1 3
        3 1 3
        4 1 2
        0 2 1
        1 2 4
        2 2 1
        3 2 4
        4 2 4
        0 3 1
        1 3 1
        2 3 1
        3 3 2
        4 3 1
      };
    \end{axis}
  \end{tikzpicture}
  \caption{Benchmark rank heat map of each policy across evaluation criteria (1 = best, 4 = worst).}
  \label{fig:policy-rank-heatmap}
\end{figure}
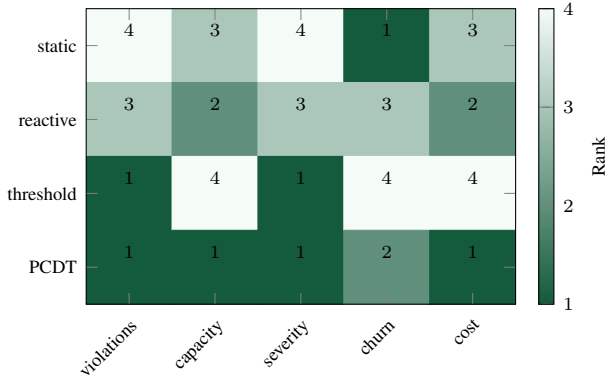

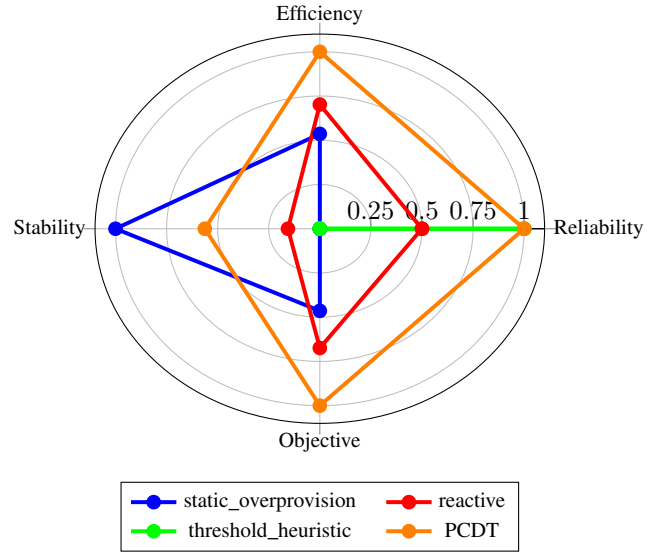
\begin{figure}[htbp]
  \centering
\begin{tikzpicture}
  \begin{polaraxis}[
    width=0.85\columnwidth, height=0.76\columnwidth,
    ytick={0.25, 0.5, 0.75, 1.0},
    xtick={0, 90, 180, 270},
    xticklabels={Reliability, Efficiency, Stability, Objective},
    xticklabel style={font=\footnotesize},
    legend style={
      at={(0.5,-0.15)},
      anchor=north,
      draw=black,
      fill=white,
      font=\footnotesize,
      legend columns=2,
      /tikz/every even column/.append style={column sep=0.3cm},
    },
  ]
    \addplot[blue, mark=*, line width=1.5pt] coordinates {(0,0.000) (90,0.536) (180,1.000) (270,0.464) (0,0.000)};
    \addlegendentry{static\_overprovision}
    \addplot[red, mark=*, line width=1.5pt] coordinates {(0,0.500) (90,0.702) (180,0.156) (270,0.674) (0,0.500)};
    \addlegendentry{reactive}
    \addplot[green, mark=*, line width=1.5pt] coordinates {(0,1.000) (90,0.000) (180,0.000) (270,0.000) (0,1.000)};
    \addlegendentry{threshold\_heuristic}
    \addplot[orange, mark=*, line width=1.5pt] coordinates {(0,1.000) (90,1.000) (180,0.563) (270,1.000) (0,1.000)};
    \addlegendentry{PCDT}
  \end{polaraxis}
\end{tikzpicture}
  \caption{Normalized multi-objective controller radar chart, outer ring is best for each metric.}
  \label{fig:policy-radar}
\end{figure}

\begin{figure}[htbp]
  \centering
\begin{tikzpicture}
  \begin{axis}[
    width=\columnwidth, height=0.57\columnwidth,
    ybar,
    bar width=0.5cm,
    ylabel={Best-fit offset (steps)},
    ylabel style={font=\small},
    xtick={0,1,2,3},
    xticklabels={static,reactive,threshold,PCDT},
    xticklabel style={font=\small},
    yticklabel style={font=\small},
    ymin=0, ymax=5,
    enlarge x limits=0.15,
    grid=major,
    grid style={line width=0.3pt, draw=gray!30},
  ]
    \addplot[fill=blue!50, draw=black] coordinates {
      (0, 0)
      (1, 1)
      (2, 1)
      (3, 4)
    };
    \node[
      draw=gray!40, fill=white, align=left, font=\scriptsize,
      anchor=north west, inner sep=4pt,
    ] at (axis cs:-0.35,4.85) {
      static: N/A \\
      reactive: lags load ($r{=}1.00$) \\
      threshold: lags load ($r{=}0.89$) \\
      PCDT: leads load ($r{=}4$)
    };
  \end{axis}
\end{tikzpicture}
  \caption{Best-fit offset (correlation lag) across compared frameworks.}
  \label{fig:lag}
\end{figure}
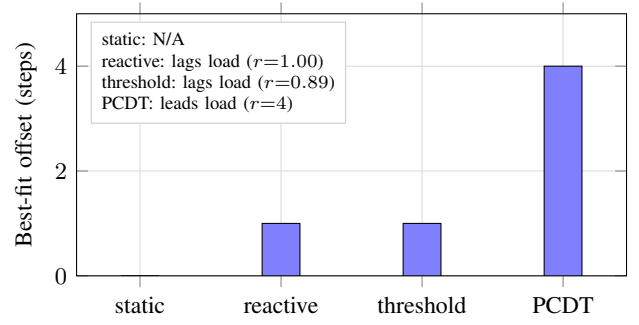

\section{Results Discussion}

\paragraph{Prediction accuracy}
PCDT achieves the lowest allocation error among all evaluated frameworks, with a MAE of approximately $0.028$ and a RMSE of approximately $0.035$, compared with $0.049$ and $0.052$, respectively, for the strongest baseline (reactive control) as depicted in Figure~\ref{fig:mae-rmse-bar}. The improved prediction accuracy shown with a lower error signified more efficient provisioning, see Figure~\ref{fig:capacity-boxplot}. PCDT allocates the lowest mean normalized capacity ($\approx0.05$), compared with approximately $0.07$--$0.10$ for the baseline controllers.

\paragraph{Operating cost}
These improvements result in the best overall operational performance. As shown in Figure~\ref{fig:policy-rank-heatmap}, PCDT ranks first in violation rate, capacity utilization, violation severity, and total cost, while Figure~\ref{fig:policy-radar} illustrates its superior trade-off between efficiency, reliability, stability, and optimization objectives.

\paragraph{Stability}
PCDT ranks second in reconfiguration churn, with static overprovisioning achieving the highest stability simply because it never changes its allocation. This stability, however, comes at the cost of significantly higher resource usage as shown in Figures~\ref{fig:policy-rank-heatmap} and~\ref{fig:policy-radar}.

\paragraph{Proactive behavior (lag analysis)}
Lag analysis explains these gains. Reactive and threshold-based controllers align with the traffic one time step \emph{after} load changes ( $\approx 1.00$ and $\approx0.89$) respectively, whereas PCDT aligns four time steps \emph{ahead} of the load, enabling lower capacity allocation without increasing SLA violations see Figure~\ref{fig:lag}. Additionally, Table~\ref{tab:standards-alignment} shows that PCDT aligns with the 6G self-optimizing network vision and the intent-driven management framework defined in 3GPP TS~28.312.

\begin{table}[t]
\caption{Alignment of PCDT with 6G Vision and 3GPP Standards \cite{sharma2023sla, 3gpp_ts28312_r19}}
\label{tab:standards-alignment}
\centering
\footnotesize
\renewcommand{\arraystretch}{1.15}
\begin{tabularx}{\columnwidth}{|l|X|X|}
\hline
\textbf{Framework} & \textbf{6G Property} & \textbf{3GPP Property} \\
\hline
PCDT &
\checkmark\ Predictive/proactive network management. Closed-loop sense$\to$predict$\to$act autonomy. &
\checkmark\ Intent-driven network management (3GPP TS~28.312) SLA/latency target expressed as controller intent. \\
\hline
\end{tabularx}
\end{table}

\section{Conclusion}
This work demonstrated that cognitive digital twins should evolve from synchronized digital replicas toward predictive decision-making entities. Rather than merely observing network state, future digital twins should proactively infer future network intent and allocate resources before service degradation occurs. PCDT has represented one practical realization of this transition toward AI-native autonomous 6G networks. Future works will focus on multiple cell analysis, integration of multiple forecasting frameworks, and evaluation on different changing traffic distributions.

\section*{Acknowledgment}
This work was partially funded by the European Union. Views and opinions expressed are however those of the author(s) only and do not necessarily reflect those of the European Union or the Research Executive Agency. Neither the European Union nor the granting authority can be held responsible for them. Grant Agreement no. 101236523 AeroNet. Additional funding was provided by the Bavarian Government through the High-Tech Agenda (HTA).

\bibliographystyle{IEEEtran}
\bibliography{references}

\end{document}